\documentclass[aps,twocolumn,10pt,amsfonts,amssymb,amsmath,preprintnumbers]{revtex4}
\usepackage[latin1]{inputenc}
\usepackage{epsfig}
\usepackage{amsfonts}
\usepackage{amssymb}
\usepackage{graphicx}
\graphicspath{{figures/}}

\begin{document}
\title{Control of coherence resonance in semiconductor superlattices}

\author{Johanne \surname{Hizanidis}}
% \email{hizanidis@itp.physik.tu-berlin.de}
\author{Eckehard \surname{Sch\"oll}}
\email{schoell@physik.tu-berlin.de}
\affiliation{Instutut f{\"u}r Theoretische Physik, Technische Universit{\"a}t Berlin, Hardenbergsta{\ss}e 36, D-10623 Berlin, Germany}

\date{\today}

 %\preprint{draft \today}

\pacs{
05.45.-a, % Nonlinear dynamics and nonlinear dynamical systems
73.21.Cd, % Superlattices
05.40.-a % Fluctuation phenomena, random processes, noise, and Brownian motion
  %05.45.Ac, % Low-dimensional chaos
  %05.45.Pq, % Numerical simulations of chaotic systems
  %02.30.Oz, % Bifurcation theory 
  % 72.20.Ht, % High-field and nonlinear effects
 72.20.Ht, 72.70.+m}

\begin{abstract}
We study the effect of time-delayed feedback control and Gaussian white noise on the
spatio-temporal charge dynamics in a semiconductor superlattice. The system is prepared 
in a regime where the deterministic dynamics is close to a global bifurcation, 
namely a saddle-node bifurcation on a limit cycle ({\it SNIPER}).  In the absence of 
control, noise can induce electron charge front motion through the entire device,
and coherence resonance is observed.
We show that with appropriate selection of the time-delayed feedback parameters the 
effect of coherence resonance can either be enhanced or destroyed, and the coherence of
stochastic domain motion at low noise intensity is dramatically increased. Additionally, 
the purely delay-induced dynamics in the system is investigated, and a homoclinic 
bifurcation of a limit cycle is found.
\end{abstract}

\maketitle

\section{Introduction}
In contrast to the problem of controlling deterministic chaos, for which
a number of methods have been proposed and successfully applied \cite{SCH07},
control of noise-induced and noise-mediated motion is a significantly less studied 
concept. Noise-induced dynamics refers to the case where there are no self-sustained 
oscillations in the absence of noise. The deterministic system rests in a stable steady 
state, e.\,g.\ a stable fixed point, and may be pushed out from it by random 
fluctuations. On the other hand, in the case of noise-mediated motion, the system 
already exhibits deterministic oscillations. Generally, the addition of noise may not 
only smear-out those oscillations but may also induce qualitatively new structures and 
dynamics, like coherence resonance \cite{HU93a,PIK97}.

The ability to control the properties of noisy oscillations (both noise-induced and noise-mediated) is very often of practical relevance. This usually implies the enhancement in the regularity of motion. However, in some cases, for instance in medical or biological applications,
too much regularity is unwanted since it may lead to damaging
consequences, e.\,g.\ epilepsy or Parkinson's disease \cite{TAS02}.

Previous works have mainly concentrated on the control of
stochastic oscillations in low-dimensional simple models
\cite{GOL03,JAN03,BAL04,POM05a,HAU06,PRA07,POT07,POT08}, 
while control of
noise-induced dynamics in spatially extended systems \cite{HIZ05,STE05a,BAL06} seems
still to be an open problem. This is the subject of the present paper. It is organized 
as follows. In Sect. II we will introduce the stochastic model of a semiconductor
superlattice. In Sect. III time-delayed feedback control is implemented. In Sect. IV
delay-induced spatio-temporal charge dynamics is studied in the absence of noise.
Sect. V deals with the time-delayed feedback control of stochastic spatio-temporal dynamics.

\section{Sequential tunneling model}

We consider a stochastic model for a superlattice \cite{HIZ06} 
which consists of epitaxial layers of two alternating semiconductor materials with 
different bandgaps, thus forming a periodic sequence of potential 
wells and barriers.
The superlattice is a prominent example of a semiconductor nanostructure device which may 
serve as a practically relevant nonlinear 
model system \cite{SCH00} and may find applications as an ultra-high frequency electronic 
oscillator \cite{KAS95,HOF96,SCH99h}.

Our model is based on sequential tunneling of electrons \cite{WAC02}. The resulting tunneling current density
$J_{m\to m+1}(F_m, n_m, n_{m+1})$ from well $m$ to well $m+1$ depends
only on the electric field $F_m$ between both wells and the electron
densities $n_m$ and $n_{m+1}$ in the respective wells (in units of
$cm^{-2}$), as given by eqs. (83), (86) in Ref. \cite{WAC02}.  

We simulate numerically a superlattice \cite{AMA04} of $N=100$ GaAs wells of width 
$w=8 \text{nm}$, and Al$_{0.3}$Ga$_{0.7}$As barriers of width $b=5 \text{nm}$,
doping density $N_D= 10^{11} \text{cm}^{-2}$ and at temperature $T=20 \text{K}$ 
\cite{parameters}.

The random fluctuations of the current densities are modelled by additive Gaussian white noise 
$\xi_{m}(t)$ with
\begin{equation}
\label{gauss}
\langle \xi_{m}(t)\rangle = 0,\quad
\langle \xi_m(t) \xi_{m'}(t')\rangle = \delta(t-t')\delta_{mm'},
\end{equation}
which yields the following Langevin equations ($m=1,...,N$):
\begin{equation}
\label{SL-equations}
e \frac{dn_m}{dt}=J_{m-1 \rightarrow m}+D \xi_{m}(t)-J_{m\rightarrow m+1}-D \xi_{m+1}(t),
\end{equation}
where $D$ is the noise intensity \cite{HIZ06}. Since the wells in our superlattice model
are considered to be weakly coupled and the tunneling times are much smaller
than the characteristic time scale of the global current
\begin{equation}
J=\frac{1}{N+1} \sum^{N}_{m=0}J_{m\rightarrow m+1},
\end{equation}
these noise sources can be treated as 
uncorrelated both in time and space.
Charge conservation is automatically guaranteed by adding a noise
term $\xi_m$ to each current density $J_{m-1\rightarrow m}$.

The electron densities and the electric fields are coupled by the discrete Poisson equation:
\begin{equation}
  \label{eq:poisson}
  \epsilon_r \epsilon_0 (F_{m} - F_{m-1}) = e(n_m -N_D)
\quad \text{for } m  = 1, \ldots N,
\end{equation}
where $\epsilon_r$ and  $\epsilon_0$ are the relative and absolute
permittivities, $e<0$ is the electron charge and
$F_0$ and $F_N$ are the fields at the emitter and collector barrier, respectively.
In the deterministic case, i.\,e.\ at $D=0$, rich dynamical scenarios can be observed 
including formation of charge accumulation and depletion fronts associated with
field domains bounded by these charge fronts, and with
stationary, periodic or even chaotic current oscillations \cite{PAT98,AMA03,AMA04}.
The two control parameters which are crucial for the dynamics, are
the applied voltage $U$ between emitter and collector, which gives rise to a
global constraint:
\begin{equation}
  \label{eq:voltage}
  U = - \sum_{m=0}^N F_m d,
\end{equation}
where $d$ is the superlattice period, and
the contact conductivity $\sigma$ \cite{XU07}. For the current densities at the emitter and collector we use Ohmic
boundary conditions, $J_{0 \to 1} = \sigma F_0 $ and $J_{N \to N + 1} = \sigma F_N \frac{n_N}{N_D}$.

\section{Time-delayed feedback}
Time-delayed feedback was originally proposed by Pyragas in the context of chaos control \cite{PYR92}. The 
goal of this method  is to stabilize unstable periodic orbits (UPOs) by adding, to a 
chaotic system, a control force in the form of the difference between a system variable at time $t$ and its delayed version at time $t-\tau$. Here $\tau$ is a delay time and $K$ is the feedback strength. 

Apart from the control of UPOs the stabilization of unstable fixed points can also be 
achieved by time-delayed feedback \cite{AHL04,HOE05}. This method proved to be very 
powerful and has been successfully applied to various physical systems since then 
\cite{SCH07}. Other variants have  been elaborated, e.\,g.\, 
extended time-delay autosynchronization (ETDAS) \cite{SOC94}, and have been applied not 
only to deterministic systems \cite{BEC02,UNK03,SCH03a,DAH07} including fixed points 
but to stochastic systems as well \cite{POM07,SCH08a}. 

Apart from the deliberate design of feedback control loops to manipulate the intrinsic 
dynamics, delay arises also naturally in many systems. Typical examples of such systems 
are lasers where the delay enters through the coupling to external cavities 
(optical feedback) \cite{HEI01a,SCH06a,FLU07} and neurons, where the signal propagation 
yields a delay time \cite{DAH08}. 

An easy way to implement control in the superlattice model is to
choose the output signal to be the total current density, $\overline J(t)$
and  simply add the control force to the external voltage $U$, i.\,e.\
\begin{equation}\label{eq:control_u}
U=U_0-K(\overline J(t)-\overline J(t-\tau)),
\end{equation}
where $U_0$ is the time-independent external voltage bias and $\overline J$ 
is the total current density, filtered by a low-pass filter, 
\begin{equation}
 \label{eq:LPF}
\overline J(t)=\alpha \int _{0}^{t}J(t')e^{-\alpha(t-t')}dt',
\end{equation}
with $\alpha$ being the cutoff frequency \cite{SCH03a}. 

Since both voltage and total current density are externally accessible 
global variables, such a control scheme is easy to implement experimentally. 
The low-pass filter in the current density was introduced initially for the effective control of chaotic motion in the superlattice \cite{SCH03a}. Due to the well-to-well hopping of the depletion and accumulation fronts, the current is modulated by fast small-amplitude oscillations. These high-frequency oscillations render the control loop unstable as $J(t)$ is fed back to the system and therefore they need to be filtered out.
Stochastic oscillations, like chaotic ones, exhibit the same high-frequency oscillations and thus we apply this filter here, too.

\section{Delay-induced dynamics}

\begin{figure}[tbp]
 \centering
 \epsfig{file=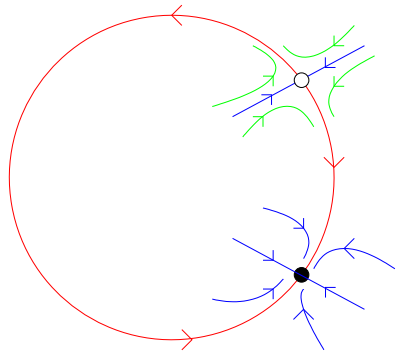,width=0.6\columnwidth} 
 \caption{(color online) Schematic phase portrait below a {\it SNIPER} bifurcation. The full circle and 
the open circle mark the stable node and the saddle-point, respectively.}
\label{fig:fig1}
 \end{figure}

In the following, the system is prepared in a stable fixed point corresponding to 
values of voltage $U=2.99V$ and conductivity $\sigma=2.082 {\Omega}^{-1}{m}^{-1}$. 
It corresponds to a stationary accumulation front separating the high-field domain attached 
to the collector from the low-field domain adjacant to the emitter. Besides the stable 
fixed point (node) there exists a saddle-point, whose unstable manifolds are connected
to the stable node, forming a closed loop in phase space as depicted schematically
in Fig.\ref{fig:fig1} \cite{HIZ06}. 
In this regime the system is very close to a saddle-node bifurcation on a limit cycle 
(saddle-node infinite period bifurcation, {\it SNIPER}). In \cite{HIZ06} a bifurcation 
analysis in the $(\sigma, U)$ plane was performed showing how {\it SNIPER} bifurcations 
govern the transition from stationary to moving field domains in the superlattice. In 
the vicinity of such a bifurcation the system is excitable \cite{LIN04} and therefore 
very sensitive to noise which is able to trigger front motion through the device. 
Moreover, the phenomenon of coherence resonance \cite{PIK97} was also confirmed in our 
model. 

In this paper, we will study how time-delayed feedback acts on the system both in the 
presence and absence of noise. when prepared in the vicinity of a {\it SNIPER} bifurcation.

\begin{figure}[tbp]
 \centering
 \epsfig{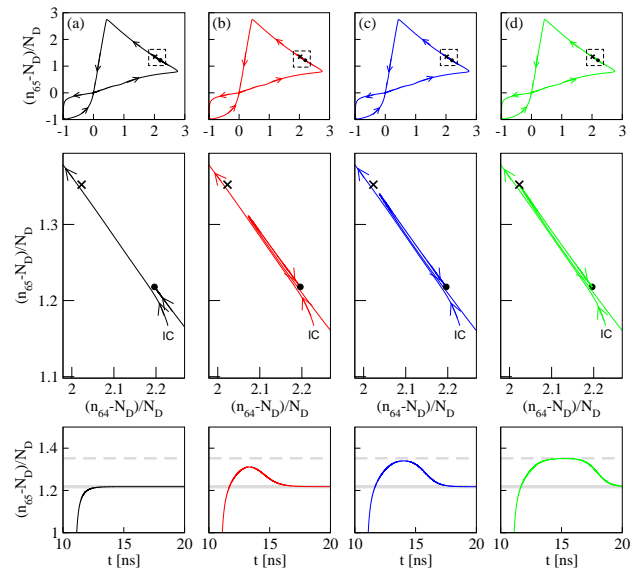}
 \caption{(color online) Top: Phase portrait in the $(n_{65},n_{64})$ plane for fixed $\tau=2ns$ and  (a) $K=0$, (b) $0.006$, (c) $0.0064$ and (d) $0.006437Vmm^2/A$.
Middle: Enlargement of corresponding phase portraits shown in upper plots.
Cross marks the position of the saddle-point, full circle marks the position of the 
stable node and $IC$ is the initial condition, which is the same in all figures. 
Bottom: Corresponding electron density time series (only the part approaching 
asymptotically the stable node  of $n_{65}$). Thick grey line denotes the value of the 
stable node and dashed grey line denotes that of the saddle-point.  $U=2.99V$ and $\sigma=2.0821 (\Omega m)^{-1}$.}
\label{fig:fig2}
 \end{figure}

In the absence of both delay and noise the only stable attractor in phase space is the 
stable node.
Regardless of the initial condition, all trajectories end there. 
In particular, when selecting the initial condition on one of the two unstable 
manifolds of the saddle-point, the system performs a long excursion along the invariant 
circle before ending in the stable node (Fig.\ref{fig:fig1}). This deterministic path is affected by the delay 
for given combinations of the two control parameters, $K$ and $\tau$. By keeping the 
delay $\tau$ fixed and considering different control amplitudes $K$ we track these orbits 
in phase space (Fig.~\ref{fig:fig2}). The top panel of Fig.~\ref{fig:fig2}(a) (K=0) shows
a trajectory which closely follows the unstable manifold of the saddle-point (cross) and
ends in the stable node (full circle), cf. blow-up in the middle panel. Note that in the
chosen 2-dimensional projection of the N-dimensional phase space the invariant circle
(saddle-node loop) is distorted to a figure-eight shape.

In Fig.~\ref{fig:fig2}(b), during the first few nanoseconds the systems acts as it would 
in the absence of delay, repelled  by the saddle-point. Control is switched on at 
$t=\tau=2ns$ when the control force begins to act. The interval $[0,2ns]$ serves as 
initial condition of the delay equation. This becomes evident in a ``twist'' in the 
trajectory just before the orbit reaches the stable node (middle panel of 
Fig.~\ref{fig:fig2}(b)).
For a moment it looks as if the system is attracted to the saddle-point instead of the 
node. This may be understood as follows: The control force shortly pulls the system off 
the phase space of the uncontrolled system pushing it towards the stable manifold of the 
saddle-point. At a critical value $K_c=0.0064375 Vmm^2/A$, the system is indeed 
``swallowed'' by the saddle and the trajectory closes in a homoclinic orbit.

In the top panel of Figs.~\ref{fig:fig2}(b)~-~\ref{fig:fig2}(d) the trajectories for 
three values of $K$ approaching this critical value are shown.
Due to the high dimensionality of the system, which is blown up to infinity due to the 
delay, the above mechanism is not clearly demonstrated in a 2-dimensional projection in 
phase space. One must zoom in carefully in order to see the deviation from the 
deterministic path due to delay (middle panels of Fig.~\ref{fig:fig2}). 
This deviation is even better visible in the bottom panels of Fig.~\ref{fig:fig2} where 
the final part of the electron density time series of $n_{65}$ is plotted. In (a) the 
deterministic trajectory is plotted and the thick grey solid and dashed lines mark the 
position of the stable node and the saddle-point, respectively. It is clear that, the 
closer one is to the homoclinic bifurcation (d), the closer to the saddle-point does the 
system reach and the longer the trajectory remains there, before ending up in the stable 
node. 

\begin{figure}[tbp]
 \centering
 \epsfig{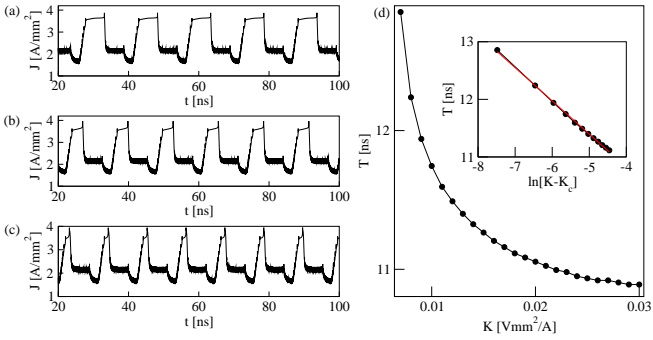}\\ 
 \caption{(a) Time series of the delay-induced limit cycle for fixed $\tau=2ns$ showing period 
lengthening as the homoclinic bifurcation is approached from above ($K>K_c$), for 
$K=0.0064379$, (b) $K=0.007$ and (c) $K=0.019Vmm^2/A$. 
(d) Period $T$ as a function of control amplitude $K$,
 and characteristic scaling law governing the homoclinic bifurcation shown in inset. $K_c=0.0064375 Vmm^2/A$, $U=2.99V$, and $\sigma=2.0821 (\Omega m)^{-1}$.}
\label{fig:fig3}
 \end{figure}

\begin{figure}[tbp]
   \centering
 \epsfig{file=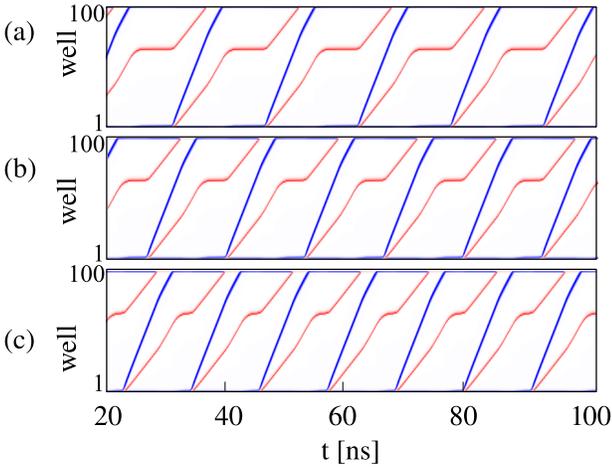,width=1.0\columnwidth}
\caption{(color online). Delay-induced front motion: Space-time plots of the electron density for 
$K=0.0064379$, (b) $K=0.007$ and (c) $K=0.019Vmm^2/A$. 
Light (red) and dark (blue) shading 
corresponds to electron accumulation and depletion fronts, respectively. 
The emitter is at the bottom, the collector at the top. Parameters as in Fig.~\ref{fig:fig3}.
} 
\label{fig:fig4}
\end{figure}

Beyond the critical value of $K_c=0.00644 Vmm^2/A$ the homoclinic orbit breaks and a limit cycle is born. 
In Figs.~\ref{fig:fig3}(a)~-~\ref{fig:fig3}(c) the time series of the current density for three different values of $K$ above $K_c$ are shown. The period clearly decreases for increasing $K$. 
Plotting it as a function of the control strength we obtain 
Fig.~\ref{fig:fig3} (d). The period $T$ shows  a characteristic scaling law \cite{KUZ95}, 
$T\sim ln[K-K_c]$, shown in the inset.
This law governs another type of global bifurcation, namely the homoclinic bifurcation.
This delay-induced dynamics is in perfect agreement with a generic two-variable model 
for the {\it SNIPER} bifurcation, which also exhibits homoclinic bifurations of a limit 
cycle if delay is added \cite{HIZ07}. 
Here, however, we have a much more complex spatio-temporal system. The corresponding 
space-time plots are shown in Fig.~\ref{fig:fig4}. It is clear that accumulation and
depletion fronts corresponding to dipole domains are created at the emitter (bottom),
and move through the device. Due to the global voltage constraint, they interact with 
the additional accumulation front in the middle of the sample, thus forming a tripole 
oscillation \cite{AMA04}.

\begin{figure}[tbp]
 \centering
 \epsfig{file=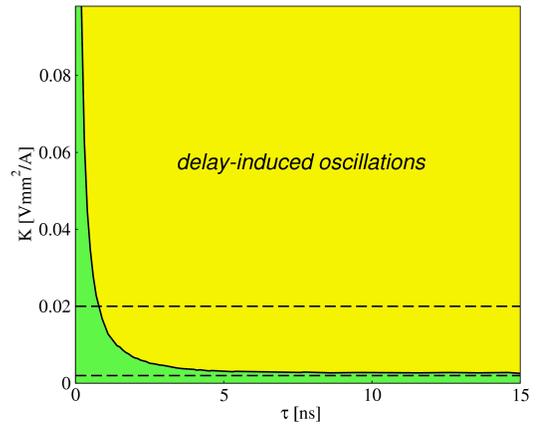,width=0.8\columnwidth}
 \caption{(color online) Bifurcation diagram in the $K$-$\tau$ plane below the {\it SNIPER} bifurcation in the superlattice. Yellow (light) area denotes the parameter regime for which delay-induced 
oscillation occur. In the green (dark) area there are no delay-induced bifurcations. The 
black line marks the delay-induced homoclinic bifurcation line.  $U=2.99V$ and $\sigma=2.0821 (\Omega m)^{-1}$.}
\label{fig:fig5}
 \end{figure}

Following the homoclinic bifurcation in the $K$-$\tau$ plane we numerically obtain the 
regime where control induces limit cycle oscillations. The result is shown in Fig.~\ref{fig:fig5}.
Crossing the bifurcation line by increasing either $K$ or $\tau$, oscillations 
are born, whose period decreases with $K$ or $\tau$. 

\section{Control of coherence resonance}
In this section, Gaussian white noise is added according to Eqs. (\ref{gauss}),
(\ref{SL-equations}). By considering two values of the 
control strength which lie inside or outside of the regime where delay induces a limit 
cycle (Fig.\ref{fig:fig5})), we will study the effect of the time delay on noise-mediated 
and noise-induced oscillations, respectively.

The regularity or coherence of noisy oscillations can be quantified by various measures. 
Here we use (i) the correlation time \cite{STR63}:
\begin{equation}
\label{eq:corrtime}
 t_{cor}=\frac{1}{\psi(0)}\int_{0}^{\infty}|\psi(s)|ds,
\end{equation}
where $\psi(s)=\langle (J(t)-\langle J \rangle)(J(t+s)-\langle J \rangle)\rangle$ is the 
autocorrelation function of the current density signal $J(t)$, and 
(ii) the normalized fluctuation of pulse durations \cite{PIK97}:
\begin{equation}
\label{eq:isi}
R_T=\frac{\sqrt{\langle T^2 \rangle-{\langle T \rangle}^2}}{\langle T \rangle},
\end{equation}
typically used for excitable systems exhibiting oscillations in the form of spike trains 
with two distinct time scales. These time scales are the 
{\em activation time}, which is the time needed to excite the system from the stable 
fixed point and the {\em excursion time} which is the time needed to return from the 
excited state to the fixed point. The sum of these two times equals the pulse duration or
period of the oscillation $T$, which denotes the time between two spikes and is also 
known as interspike interval.

\begin{figure}[tbp]
 \centering
 \epsfig{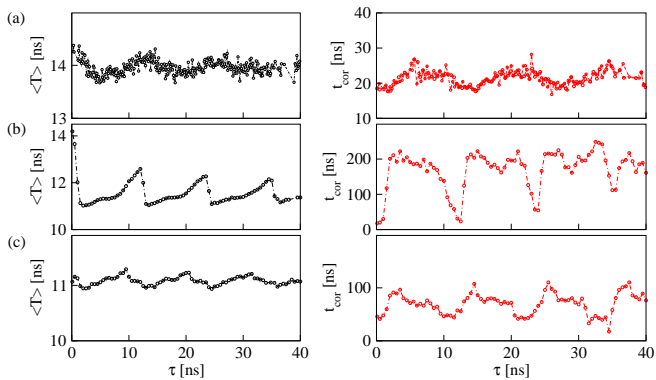} 
 \caption{(color online) Mean interspike interval $\langle T \rangle$ (left) and 
correlation time $t_{cor}$ (right) in dependence on the time delay $\tau$. 
(a) Control strength $K=0.002Vmm^2/A$ and noise intensity $D=1.0As^{1/2}/m^2$, 
(b) $K=0.02Vmm^2/A$ and $D=1.0As^{1/2}/m^2$ and (c) $K=0.02Vmm^2/A$ and $D=2.5As^{1/2}/m^2$. 
Averages over 30 time series realizations of length $T=1600ns$ have been used for the 
calculation of $t_{cor}$ and averages over 1000 periods for $\langle T \rangle$. 
$U=2.99V$ and $\sigma=2.0821 (\Omega m)^{-1}$.}
\label{fig:fig6}
 \end{figure}

By keeping the noise intensity fixed at $D=1.0As^{1/2}/m^2$ we first select a value $K=0.002Vmm^2/A$ outside the delay-induced limit cycle regime. This corresponds to the lower dashed horizontal line in Fig.~\ref{fig:fig5}. 
In the right panel of Fig.~\ref{fig:fig6}(a), the correlation time is plotted versus the 
time delay. It exhibits a slight modulation with a period close to the period of the 
noise-induced oscillations \cite{HIZ06}, $\langle T \rangle=14.5ns$  and reaches minimum 
values for $\tau=n \langle T \rangle, n \in \mathbb N$.
Overall, however, it remains close to the control-free value, $t_{cor}=19.76ns$. 
At $K=0.02Vmm^2/A$ inside the delay-induced limit cycle regime (upper dashed line in 
Fig.~\ref{fig:fig5}), this modulation is much stronger and has a period close to the delay-induced period 
($T=11ns$, see Fig.~\ref{fig:fig3}(d)). In addition, one can better distinguish between 
non-optimal and optimal values of $\tau$ at which the correlation time attains maximum 
values. This is shown in the right panel of Fig.~\ref{fig:fig6} (b). For a higher noise 
intensity (Fig.~\ref{fig:fig6} (c), right panel) the effect is similar but weaker.

Next we are interested in how the time scales are affected by the delay. We express the 
time scales through the mean interspike interval $\langle T \rangle$ and 
look at its dependence upon the time delay $\tau$ for a fixed value of the noise 
intensity, $D=1.0As^{1/2}/m^2$, and control strength $K=0.002Vmm^2/A$, chosen outside of 
the delay-induced oscillations regime (lower dashed line in Fig.~\ref{fig:fig5}). As 
shown in the left panel of Fig.~\ref{fig:fig6} (a), $\langle T \rangle$ is slightly 
modulated due to the delay with a period close to the noise-induced mean period 
($\langle T \rangle \approx 14.5ns$) \cite{HIZ06}.

In the left panel of Fig.~\ref{fig:fig6} (b) a value of $K$ inside the delay-induced 
oscillations regime is used, $K=0.02Vmm^2/A$ (upper dashed line in Fig.~\ref{fig:fig5}). 
For $\tau=0$, the mean interspike interval is equal to the noise-induced period, 
$\langle T \rangle \approx 14.5ns$ \cite{HIZ06}. As the time delay increases, 
and the delay-induced bifurcation line is crossed,
$\langle T \rangle$ sharply drops to the value of $11ns$ which corresponds to the period 
induced by the delay (see Fig.~\ref{fig:fig3}(d)). By further increase of $\tau$, 
$\langle T \rangle$ rises a little above $12ns$. Then, for $\tau=11ns$ the mean 
interspike interval decreases again and the same scenario is repeated with a modulation 
period very close to the delay-induced period.
\begin{figure}[h!]
 \centering
 \epsfig{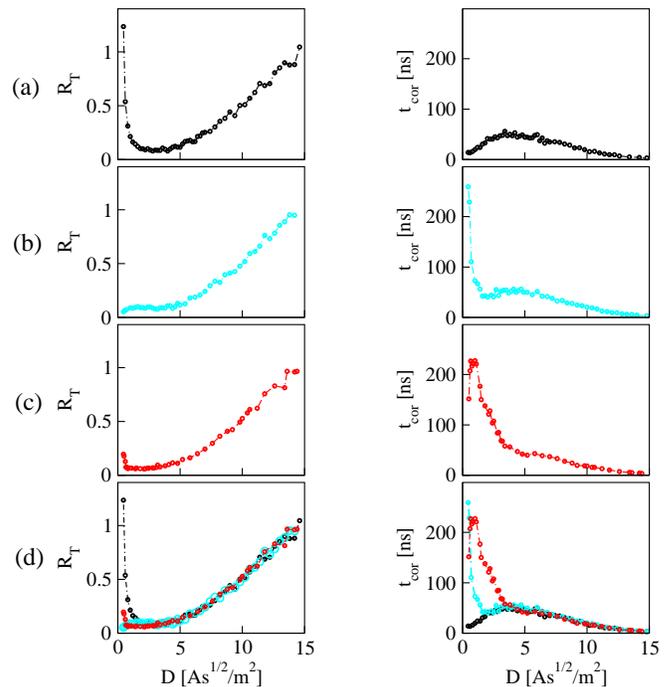}
 \caption{(color online) Correlation time (right) and normalized fluctuation of pulse 
durations (left) as a function of the noise intensity for (a)
$K=0$, (b) $(K,\tau)=(0.02Vmm^2/A,11ns)$ and (c) $(K,\tau)=(0.02Vmm^2/A,14.5ns)$. 
All three curves plotted together in (d). Averages over 30 time series realizations 
of length $T=1600ns$ have been used for the calculation of $t_{cor}$ and averages 
over 1000 periods for $R_T$. $U=2.99V$ and $\sigma=2.0821 (\Omega m)^{-1}$.}
\label{fig:fig7}
 \end{figure}
There is some resemblance to the piecewise linear dependence of $\langle T \rangle$ upon 
$\tau$ reported in other excitable systems: The FitzHugh-Nagumo model in 
\cite{JAN03,BAL04,PRA07} and the Oregonator model of the Belousov-Zhabotinsky reaction 
(under correlated noise and nonlinear delayed feedback) in \cite{BAL06} which, like our 
system, is also spatially extended. The difference to our present analysis is that in 
those models the case of a delay-induced limit cycle was excluded.
An explanation for the entrainment of the time scales by the delayed feedback 
in case of systems below a Hopf bifurcation \cite{JAN03,BAL04,POM05a,STE05a}
was given on the basis of a linear stability analysis. It was shown that the basic 
period is proportional to the inverse of the imaginary part of the eigenvalue of the 
fixed point which itself depends linearly upon $\tau$, for large time delays. The effect 
of noise and delay in excitable systems was also studied analytically in \cite{PRA07,POT08} 
based on waiting time distributions and renewal theory.

Finally we look into the dependence of the correlation time $t_{cor}$ and the normalized 
fluctuation of the interspike intervals $R_T$ on the noise intensity.  
We keep the control strength fixed to the value corresponding to Fig.~\ref{fig:fig6}(b) 
(right panel), from which we also select an optimal and a non-optimal value of the time 
delay and compare the results to the uncontrolled system.
In the left and right panel of Fig.~\ref{fig:fig7}, $R_T$ and $t_{cor}$ are plotted, 
respectively. The case $K=0$ is shown in Fig.~\ref{fig:fig7}(a) for direct comparison.
Coherence resonance shows up as a minimum of $R_T$ and a maximum of $t_{cor}$, respectively. 
For both non-optimal (Fig.~\ref{fig:fig7}(b)) and optimal $\tau$ (Fig.~\ref{fig:fig7}(c)),
there is an enhancement in the coherence at low noise intensity. Correlation times attain 
much higher values than those of the uncontrolled system, especially at low noise level.  
Similarly, the interspike interval fluctuation $R_T$ is much smaller. In addition, for 
non-optimal delay time $\tau \approx 11ns $, the effect of coherence resonance is 
suppressed (Fig.~\ref{fig:fig7}(b)). The correlation time still shows a small local 
maximum exactly where the uncontrolled system does, but for small noise intensities
the correlation time dramatically increases in a monotonic way to much larger values of
$t_{cor}$. On the other hand, for optimal $\tau \approx 14.5ns$ (Fig.~\ref{fig:fig7}(c)), 
coherence resonance is maintained and both $t_{cor}$ and $R_T$ show a maximum and 
minimum, respectively, but at a much lower noise intensity than in the free system. 
The comparison between all three cases is better visible in Fig.~\ref{fig:fig7}(d)  
where the three curves are plotted together.

\section{Conclusions}
We have shown that by applying a time-delayed feedback force to a semiconductor 
superlattice stationary field domains (bounded by charge accumulation and depletion fronts)
can be transformed into travelling domains in a homoclinic bifurcation of a limit cycle if 
the system is prepared below a saddle-node infinite period {\it SNIPER} bifurcation. 
With the addition of Gaussian white noise, control results in a modulation 
of both the coherence and time scales of the system with the time delay.
The periodicity of this  modulation is determined by the competition between the 
different time scales imposed by noise and control and their dependence on the noise 
intensity and time-delay, respectively.  We distinguish between optimal and non-optimal 
time delays at which the coherenc resonance effect is enhanced or destroyed, respectively.
In both cases the correlation times of stochastic domain motion are dramatically increased
at low noise intensities.

\section{Acknowledgements}
This work was supported by DFG in the framework of Sfb 555. 

%\bibliographystyle{prsty-fullauthor}
%\bibliography{ref}

\end{document}